\documentclass{article}
\usepackage{icproc}
\usepackage{amsfonts,euscript}

\begin{document}
\newcommand{\W}{{\mathsf W}}
\def\ope[#1][#2]{\mathord{#2\over{\ifnum#1=1 {z-w} \else {(z-w)^{#1}}\fi}}}
\newcommand{\reg}{\text{reg}}
\newcommand{\tG}{{\ensuremath{\mathbb G}}}
\newcommand{\tJ}{{\ensuremath{\mathbb J}}}
\newcommand{\tT}{{\ensuremath{\mathbb T}}}
\newcommand{\tF}{{\ensuremath{\mathbb F}}}
\renewcommand{\d}{\partial}
\newcommand{\tfrac}[2]{\textstyle \frac{#1}{#2}}
\newcommand{\half}{\tfrac{1}{2}}

\talk{UNTWISTING TOPOLOGICAL FIELD THEORIES}

\author{JM FIGUEROA-O'FARRILL}

\address{Department of Physics, Queen Mary and Westfield College,\\
LONDON E1 4NS, UK}

\maketitle
\abstracts{A method is presented by which a hidden $N{=}2$
superconformal symmetry can be exhibited in a string theory or indeed
in a topological conformal field theory.  More precisely, we present
strong evidence, based on calculations with string theories, in favour
of the conjecture that any topological conformal field theory can be
obtained by twisting an $N{=}2$ superconformal field theory.  (Talk
given at the Workshop on Gauge Theories, Applied Supersymmetry and
Quantum Gravity held at Imperial College, London, 5-10 July 1996.)}

\section{Motivation}

Generic string theories are theories of two-dimensional quantum
gravity coupled to conformal matter.  Since two-dimensional gravity
has no propagating degrees of freedom, it is not surprising that one
can make progress in its study by studying two dimensional topological
quantum field theories.  The study of these theories in turn benefits
from the study of those theories which in addition possess conformal
invariance; since just as in the non-topological theories, topological
conformal field theories (TCFTs) can be deformed to study the space of
topological field theories.  A large class of topological conformal
field theories can be constructed starting from any $N{=}2$
superconformal field theory by the twisting procedure of
Witten~\cite{Witten} and Eguchi--Yang~\cite{EYtwist}.  An important goal
in the study of a TCFT is to compute the spectrum of physical states
and determine its algebraic structure.  For those TCFTs which can be
obtained by twisting, our increasing knowledge of the representation
theory of the $N{=}2$ superconformal algebra can be brought to bear on
this problem.  For example, there is an intimate
relation~\cite{Aliosha} between singular vectors of the $N{=}2$
superconformal algebra and the BRST cohomology of the TCFT.

At the same time, seemingly different TCFTs are sometimes found to
have the same physical spectrum -- e.g., $c{=}1$ string and
$SL(2)/SL(2)$ gauged WZW model~\cite{Aharony}.  Such isomorphisms are
often {\em a posteriori\/} results; that is, {\em after\/} computing
the physical spectrum in both theories.  Having a fixed algebraic
structure as reference {\em before\/} computing cohomology, makes the
comparison of theories easier, especially in those cases where a
complicated matter sector makes the computation of the BRST cohomology
practically impossible.  This common reference structure is a
(twisted) $N{=}2$ superconformal algebra (SCA).~\cite{Sadov}

In summary, a hidden $N{=}2$ superconformal symmetry is a desirable
feature of a TCFT either for the computation of the spectrum or to
show the equivalence between theories.  In this talk we will see how
to go about constructing them.  {\em A word of caution, however\/}.
It is well-known~\cite{BeatrizAliosha,Bershadsky} that many
noncritical string theories possess a hidden $N{=}2$ superconformal
symmetry.  The $N{=}2$ symmetry that we will describe in this talk
does {\em not\/} agree with this one.  In a sense that we will make
more precise below, the $N{=}2$ superconformal symmetry of the
noncritical strings does not preserve the natural grading in the
spectrum of the TCFT, whereas by construction the $N{=}2$
superconformal symmetry we will exhibit in virtually any string theory
does.

\section{Topological conformal field theories}

Roughly, a TCFT is a conformal field theory (CFT) with a BRST
symmetry: $Q^2 = 0$, and where the energy-momentum tensor $T$ is BRST
invariant: $T = [Q,X]$ for some $X$.  Since $T$ generates
translations, correlation functions of BRST invariant fields are
locally constant and therefore topological in nature.  By definition,
the physical spectrum of the TCFT is the BRST cohomology
$H^\bullet(Q)$.  The BRST cohomology inherits many operations from the
underlying conformal field theory.  First of all it is a graded
commutative associative algebra (the ``ground ring''~\cite{WittenGR});
but in addition, in all known examples, it has the structure of a
Batalin--Vilkovisky algebra (see below).

\subsection{Twisted $N{=}2$ superconformal algebras}

The canonical example of a TCFT is any CFT possessing (twisted)
$N{=}2$ superconformal invariance.  In other words, a CFT which
affords a realisation of the following algebra with generators $\tJ$,
$\tG^{\pm}$ and $\tT$:

\begin{eqnarray*}
\tG^\pm(z) \tG^\pm(w) &=& 0\\
\tG^+(z) \tG^-(w) &=& \ope[3][d] + \ope[2][\tJ(w)] + 
       \ope[1][\tT(w)] \\
\tJ(z) \tG^\pm(w) &=& \ope[1][\pm\tG^\pm(w)]~,
\end{eqnarray*}
with all other operator product expansions following
from these by associativity~\cite{FN=2,GetzlerMT}.

For such a TCFT the BRST operator $Q$ can be identified with the zero
mode of $\tG^+$, whence the energy-momentum tensor obeys $\tT =
[Q,\tG^-]$ and the BRST cohomology agrees with the chiral ring.  The
BRST cohomology $H \equiv H^\bullet(Q)$ admits a nontrivial algebraic
structure:
\begin{itemize}
\item $H = \bigoplus_n H^n$ is graded by the zero mode of $\tJ$;
\item $H$ has a commutative associative multiplication $\circ: H^p
\otimes H^q \to H^{p+q}$ induced from the normal-ordered product; and
\item $H$ has a linear map $\Delta: H^p \to H^{p-1}$ obeying $\Delta^2
= 0$, where $\Delta$ is given by the zero mode of $\tG^-$.
\end{itemize}
The resulting algebraic structure $(H^\bullet, \circ, \Delta)$ is
called a {\em Batalin-Vilkovisky (BV)
algebra\/}~\cite{LZ,Getzler,PS,Akman,Huang,KSV}.

\subsection{Other topological conformal field theories}

Other topological field theories are known which do not obviously come
from twisting an $N{=}2$ superconformal algebra.  For example, the
hidden symmetry of the $G/G$ gauged WZW model~\cite{Ramallo,FSN=0} is
the Kazama algebra~\cite{Kazama}, a nontrivial generalisation of the
$N{=}2$ superconformal algebra.  It has generators $\tJ$, $\tG^{\pm}$,
$\tT$, $\tF$, $\Phi$, subject to the operator product expansions:
\begin{eqnarray*}
\tG^+(z) \tG^+(w) &=& 0\\
\tG^+(z) \tG^-(w) &=& \ope[3][d] + \ope[2][\tJ(w)] + 
       \ope[1][\tT(w)]\\
\tJ(z) \tG^\pm(w) &=& \ope[1][\pm\tG^\pm(w)]\\
\tG^-(z) \tG^-(w) &=& \ope[1][-2\tF(w)]\\
\tG^+(z) \Phi(w) &=& \ope[1][\tF(w)] \\
\tJ(z) \Phi(w) &=& \ope[1][-3\Phi(w)]~.
\end{eqnarray*}
As for the $N{=}2$ superconformal algebra, the other OPEs follow from
these by associativity~\cite{GetzlerMP}.  We have used this notation
for the generators of the Kazama algebra to emphasise the similarities
and the differences with the $N{=}2$ superconformal algebra: setting
$\tF$ and $\Phi$ to zero recovers the $N{=}2$ superconformal algebra.
In particular the BRST operator $Q$ is again the zero mode of $\tG^+$
and the energy-momentum tensor $\tT$ is again BRST-exact.  Despite
the fact that the $G/G$ TCFT comes from twisting a Kazama algebra
instead of an $N{=}2$ superconformal algebra, the physical spectrum
$H^\bullet(Q)$ is again a BV algebra~\cite{GetzlerMP,FSN=0}.

String theories constitute an important class of TCFTs whose
underlying symmetry is neither a twisted Kazama algebra or a twisted
$N{=}2$ SCA.  In any string theory we can define generators $\tJ$,
$\tG^\pm$, and $\tT$ by:
\begin{eqnarray*}
\tG^+ &=& j_{\mathrm{BRST}} + \partial(\cdots)\\
\tG^- &=& b\\
\tT &=& T_{\mathrm{matter}} + T_{\mathrm{Liouville}} +
T_{\mathrm{ghosts}}\\
\tJ &=& J_{\mathrm{ghosts}} + \cdots
\end{eqnarray*}
These fields will generally fail to close and the algebra which they
generate will contain further fields.  Nevertheless all string
theories, regardless of the background, give rise to a BV algebra in
cohomology~\cite{WittenGR,WZ,LZ,BMP}.  Those backgrounds where the
Liouville terms are present (i.e., those corresponding to noncritical
strings) do possess a hidden $N{=}2$ superconformal
algebra~\cite{BeatrizAliosha,Bershadsky}, but the expression for $\tJ$
is such that it receives a nontrivial contribution from the Liouville
momentum.  Therefore the grading is no longer by ghost number and
hence the chiral ring of the $N{=}2$ superconformal algebra and the
BRST cohomology of the noncritical string are {\em not\/} isomorphic
as graded algebras, although they are isomorphic if we do forget about
the grading.

\subsection{Topological conformal algebras}

All these examples of TCFTs share some common properties which are
encapsulated in the notion of a {\em topological conformal algebra} or
TCA.

A TCA is generated by fields $\tJ$, $\tG^{\pm}$, $\tT$, \ldots
subject to the following conditions:
\begin{itemize}
\item the zero mode $Q$ of $\tG^+$ obeys $Q^2 = 0$ and $\tT = [Q,
\tG^-]$;
\item the zero mode of $\tJ$ provides a grading relative to which
$\tG^\pm$ has degree $\pm 1$;
\item the zero mode of $\tG^-$ defines an operation $\Delta$ in
cohomology which obeys $\Delta^2 = 0$.
\end{itemize}
With these properties it can be shown that the cohomology
$H^\bullet(Q)$ is a BV algebra~\cite{LZ,Getzler,PS,Akman,Huang,KSV}.

Let us say that two TCFTs are {\em (cohomologically) equivalent\/} if
their BRST cohomologies are isomorphic {\em as BV algebras\/}.

By {\em untwisting\/} a TCFT we mean showing that it is equivalent (in
the above sense) to one obtained by twisting a $N{=}2$ SCA.  In the
next section we show how to untwist a large class of TCFTs.

\section{Untwisting topological conformal field theories}

Our general strategy to untwist a TCFT will be the following:
\begin{itemize}
\item add new degrees of freedom without changing the physical
spectrum; and 
\item use these new degrees of freedom to redefine the currents in
order to obey a twisted $N{=}2$ SCA.
\end{itemize}

In order to accomplish the first point, we need to discuss a very
important kind of TCFT---a {\em trivial\/} TCFT, denoted ${\EuScript
T}_{\EuScript K}$.  This trivial TCFT is constructed out of a pair of
BC systems: one fermionic $(b,c)$ and one bosonic $(\beta,\gamma)$
with the usual operator product expansions, and is generated by the
following fields:
\begin{eqnarray*}
\tG^+_{\EuScript K} &=& b\gamma\\
\tG^-_{\EuScript K} &=& \lambda \partial c\beta + (\lambda -1)
c\partial\beta\\ 
\tJ_{\EuScript K} &=& (1-\lambda) bc + \lambda \beta\gamma\\ 
\tT_{\EuScript K} &=& \lambda \left( \beta\partial\gamma - b\partial
c\right) + (\lambda -1) \left( \partial\beta\gamma - \partial b c
\right)~.
\end{eqnarray*}
For any $\lambda$, the above generators obey a twisted $N{=}2$ SCA.
(For $\lambda=2$, one can also add to $\tG^-$ a term proportional to
$b$.)  The chiral ring couldn't be simpler:
\begin{displaymath}
H^n(Q) = \left\{ \begin{array}{ll}
                \langle 1 \rangle & \mbox{for $n=0$}\\
                0 & \mbox{otherwise.}
		\end{array}
          \right.
\end{displaymath}

Now let $\tJ_0$, $\tG^{\pm}_0$, $\tT_0$,... generate a TCA ${\EuScript
T}_0$.  Define new generators:
\begin{eqnarray*}
\tG^+ &=& \tG^+_0 + \tG^+_{\EuScript K}\\
\tG^- &=& \tG^-_0 + \tG^-_{\EuScript K}\\
\tT &=& \tT_0 + \tT_{\EuScript K}\\
\tJ &=& \tJ_0 + \tJ_{\EuScript K}~.
\end{eqnarray*}
Using the K\"unneth theorem and the triviality of the the TCA
${\EuScript T}_{\EuScript K}$, it is easy to see that $\tJ$,
$\tG^{\pm}$, $\tT$,... generate a TCA ${\EuScript T}$, which is
(cohomologically) equivalent to ${\EuScript T}_0$.  Moreover, in every
example we have tried, one can further deform $\tJ$, $\tG^{\pm}$,
$\tT$ in such a way that the new deformed $\tJ$, $\tG^{\pm}$, $\tT$
obey a twisted $N{=}2$ SCA. Furthermore, the cohomology of the
deformed BRST operator (the zero mode of the new $\tG^+$) is
isomorphic as a BV algebra to the original BRST cohomology.

Let us now look at some examples.

\subsection{The bosonic string}
Consider a generic bosonic string background given by a CFT with
$c{=}26$ and energy-momentum tensor $T$.  The following generators
define an untwisting of the TCFT defined by the string theory:
\begin{eqnarray*}
\tG^+ &=& T\tilde c + \tilde b \tilde c\d \tilde c + \tfrac{3}{2} \d^2
\tilde c + \tG^+_{\EuScript K} + \d X\\
\tG^- &=& \tilde b  + \tG^-_{\EuScript K}\\
\tJ &=& - \tilde b\tilde c + \tJ_{\EuScript K} + (bc -
\beta\gamma) + \d Y\\
\tT &=& T - 2 \tilde b\d \tilde c - \d \tilde b \tilde c 
+ \tT_{\EuScript K}~,
\end{eqnarray*}
where
\begin{eqnarray*}
X &=& - \tilde c(bc - \beta\gamma) - \beta c\tilde c\d\tilde c\\
Y &=& \tilde c c \beta~.
\end{eqnarray*}

This result was first obtained~\cite{TriesteTalk} using the
Berkovits-Vafa embedding of the bosonic string in the NSR
string~\cite{BerkVafa}, where $(b,c)$ play the role of the
anticommuting fields defined by Berkovits-Vafa to embed the string and
$(\beta,\gamma)$ are the superconformal ghosts needed in the NSR
string.  In this case the parameter $\lambda$ in the trivial TCA
${\EuScript T}_{\EuScript K}$ was set equal to $\tfrac{3}{2}$;
although in the above equation $\lambda$ is free.

\subsection{The NSR string}
Consider a generic NSR string background given by a SCFT with
$c{=}15$, energy-momentum tensor $T$ and supercurrent $G$.
A possible untwisting is given by:
\begin{eqnarray*}
\tG^+ &=& T \tilde c + G \tilde\gamma + \tilde b \tilde c\d \tilde c -
\tilde b \tilde\gamma^2 + \tilde\beta\tilde\gamma\tilde c - \half
\tilde\beta\tilde\gamma\d\tilde c\\
&& + \half \d^2\tilde c - \half \d\left(\tilde\beta\tilde\gamma\tilde
c\right) + \tG^+_{\EuScript K} + \d X\\
\tG^- &=& \tilde b + \tG^-_{\EuScript K}\\
\tJ &=& - \tilde b\tilde c + \tilde\beta \tilde\gamma +
\tJ_{\EuScript K} + \tfrac{1}{2}(bc - \beta\gamma) + \d Y\\
\tT &=& T - 2 \tilde b\d \tilde c - \d \tilde b \tilde c 
+ \tfrac{3}{2} \tilde\beta\d\tilde\gamma + \half
\d\tilde\beta\tilde\gamma + \tT_{\EuScript K}
\end{eqnarray*}
where
\begin{eqnarray*}
X &=& - \tilde c(bc - \beta\gamma) - \beta c(\tilde c\d\tilde c -
\tilde\gamma^2)\\
Y &=& \half \tilde c c \beta~.
\end{eqnarray*}
That any NSR string is (cohomologically) equivalent to an $N{=}2$
SCA, had already been noted by Marcus~\cite{Marcus} using the
embedding of NSR string into the $N{=}2$ string discovered by
Berkovits and Vafa.

\subsection{Other (string) theories}

Similar results hold for the Kazama algebra, a result independently
obtained by Getzler~\cite{GetzlerMP}.  A similar but even simpler
result holds for the $N{=}2$ string~\cite{Joaquim,GiveonRocek}, where
the $(b,c,\beta,\gamma)$ system is not necessary, since the $U(1)$
current $\tJ$ is null in this case.  The case for $\W$-strings is
similar.  We have shown~\cite{JMFPaper} that any $\W_3$-string can be
untwisted, although the expressions are a little messier than for the
above strings.  Other $\W$-string theories are known to exist (e.g.,
$\W_4$) and we are confident that the method generalises, although the
details are bound to get messier still.

Notice however that if, as Drinfel'd--Sokolov reduction suggests,
string theories are (cohomologically) equivalent to $G/G$ gauged WZW
models, then this conjecture would follow from the fact that $G/G$
gauged WZW models are described by a Kazama algebra.  Of course, we
are still far from being able to fully exploit this, since the quantum
BRST operators for most $\W$-string theories only exist at a
conjectural level.

\section{Conclusions}

We have seen that many TCFTs are (cohomologically) equivalent to
twisted $N{=}2$ SCFTs.  The obvious conjecture is that this is the
case for {\em all\/} TCFTs.  No counterexample is known, but not all
known TCFTs have been checked (e.g., $\W_4$-strings, generalised
$\W$-strings).  However the complexity of these remaining cases begs
for a more conceptual proof, for which we would first need to
understand the nature of the obstruction to the ``$N{=}2$-ness'' of a
TCFT.  We are in the curious predicament that we know in practice how
to kill the obstruction, without a deep understanding of it.

\section*{Acknowledgements}
It is a pleasure to thank the organisers for the opportunity to talk
at this conference, and in particular Kelly Stelle for arranging a
most memorable banquet, and Kris Thielemans for his part in choosing
the menu. ({\em Yum!\/})  The work described in this talk was funded
in part by the EPSRC under contract GR/K57824.

\section*{References}
\newcommand{\NPB}[3]{{\sl Nucl. Phys.} {\bf B#1} (#2) #3}
\newcommand{\CMP}[3]{{\sl Comm. Math. Phys.} {\bf #1} (#2) #3}
\newcommand{\PLB}[3]{{\sl Phys. Lett.} {\bf #1B} (#2) #3}
\newcommand{\MPLA}[3]{{\sl Mod. Phys. Lett.} {\bf A#1} (#2) #3}
\newcommand{\PRL}[3]{{\sl Phys. Rev. Lett.} {\bf #1} (#2) #3}


\begin{thebibliography}{99}

\bibitem{Aharony}
O~Aharony, O~Ganor, J~Sonnenschein and S~Yankielowicz,
{\em $c{=}1$ string theory as a topological $G/G$ model\/},
{\tt hep-th/9302027}, \PLB{305}{1993}{35-42}.

\bibitem{Akman}
F~Akman, {\em On some Generalizations of
Batalin--Vilkovisky algebras}, {\tt q-alg/9506027}.

\bibitem{BerkVafa}
N~Berkovits and C~Vafa, {\em On the Uniqueness of
String Theory}, {\tt hep-th/9310170}, \MPLA{9}{1994}{653-664}.

\bibitem{Bershadsky}
M~Bershadsky, W~Lerche, D~Nemeschansky, and
N~Warner, {\em Extended $N{=}2$ superconformal structure of gravity
and $\W$-gravity coupled to matter}, {\tt hep-th/9211040},
\NPB{401}{1993}{304-347}.

\bibitem{BMP}
P~Bouwknegt, J~McCarthy and K~Pilch, {\em The $\W_3$ algebra: modules,
semi-infinite cohomology and BV-algebras\/}, {\tt hep-th/9509119}; and
{\em BV-structure of the cohomology of nilpotent subalgebras and the
geometry of ($W$-)strings\/}, {\tt hep-th/9512032}.

\bibitem{EYtwist}
T~Eguchi and SK~Yang, {\em $N{=}2$ superconformal
models as topological field theories}, \MPLA{5}{1990}{1693-1701}.

\bibitem{FN=2}
JM~Figueroa-O'Farrill, {\em Affine algebras, $N{=}2$ superconformal
algebras, and gauged WZNW models}, {\tt hep-th/9306164},
\PLB{316}{1993}{496}.

\bibitem{N=0inN=1}
JM~Figueroa-O'Farrill, {\em On the Universal
String Theory}, {\tt hep-th/9310200}, \PLB{321}{1994}{344-348};\\
H~Ishikawa and M~Kato, {\em Note on $N{=}0$ string
as $N{=}1$ string}, {\tt hep-th/9311139}, \MPLA{9}{1994}{725-728}.

\bibitem{FSN=0}
JM~Figueroa-O'Farrill and S~Stanciu, {\em Nonreductive
WZW models and their CFTs}, {\tt hep-th/9506151}.

\bibitem{TriesteTalk}
JM~Figueroa-O'Farrill, {\em Are all TCFTs obtained by
twisting $N{=}2$ TCFTs?}, {\tt hep-th/9507024}.

\bibitem{JMFPaper}
JM~Figueroa-O'Farrill, {\em $N{=}2$ structures in string theories\/},
{\tt hep-th/9507145}.

\bibitem{BeatrizAliosha}
B~Gato-Rivera and AM~Semikhatov, {\em $d\leq 1
\cup d\geq 25$ and $\W$ constraints from BRST invariance in the $c\neq
3$ topological algebra}, {\tt hep-th/9207004}, \PLB{293}{1992}{72}.

\bibitem{Getzler}
E~Getzler, {\em Batalin--Vilkovisky algebras and
two-dimensional topological field theories}, {\tt hep-th/9212043}, 
\CMP{159}{1994}{265-285}.

\bibitem{GetzlerMT}
E~Getzler, {\em Manin triples and $N{=}2$ superconformal field
theory}, {\tt hep-th/9307041}.

\bibitem{GetzlerMP}
E~Getzler, {\em Manin pairs and topological field theory},
{\tt hep-th/9309057}.

\bibitem{GiveonRocek}
A~Giveon and M~Ro\v cek, {\em On the BRST Operator
Structure of the $N{=}2$ String}, {\tt hep-th/9302049},
\NPB{400}{1993}{145-160}.

\bibitem{Joaquim}
J~Gomis and H~Suzuki, {\em $N{=}2$ string as a
topological conformal algebra}, {\tt hep-th/9111059},
\PLB{278}{1992}{266-270}.

\bibitem{Huang}
YZ~Huang, {\em Operadic formulation of topological
vertex algebras and Gerstenhaber or Batalin--Vilkovisky algebras},
{\tt hep-th/9306021}, \CMP{164}{1994}{105-144}.

\bibitem{Ramallo}
JM~Isidro and AV~Ramallo, {\em Topological current
algebras in two-dimensions}, {\tt hep-th/9307176},
\PLB{316}{1993}{488-495}.

\bibitem{Kazama}
Y~Kazama, {\em Novel topological field theories},
\MPLA{6}{1991}{1321-1332}.

\bibitem{KSV}
T~Kimura, J~Stasheff, and AA~Voronov, {\em On operad
structures of moduli spaces and string theory}, {\tt hep-th/9307114}.

\bibitem{LZ}
BH~Lian and GJ~Zuckerman, {\em New perspectives on the
BRST-algebraic structure of string theory}, {\tt hep-th/9211072},
\CMP{154}{1993}{613-646}; {\em Some classical and quantum algebras},
{\tt hep-th/9404010}; {\em Algebraic and geometric structures in string
backgrounds}, {\tt hep-th/9506210}.

\bibitem{Marcus}
N~Marcus, {\em The $N{=}1$ superstring as a
topological field theory},\\
{\tt hep-th/9405039}, \PRL{73}{1994}{1071-1074}.

\bibitem{PS}
M~Penkava and A~Schwarz, {\em On some algebraic structures
arising in string theory}, {\tt hep-th/9212072}.

\bibitem{Sadov}
V~Sadov, {\em The hamiltonian reduction of the BRST complex and
$N{=}2$ SUSY\/}, {\tt hep-th/9304049}.

\bibitem{Aliosha}
AM~Semikhatov, {\em Verma modules, extremal vectors,
and singular vectors on the non-critical $N{=}2$ string worldsheet},
{\tt hep-th/9610084};\\
AM~Semikhatov and IY~Tipunin, {\em All
singular vectors of the $N{=}2$ superconformal algebra via the
algebraic continuation approach}, {\tt hep-th/9604176}; and {\em
$sl(2)_{-4}$ WZW model as an $N{=}4$ supersymmetric bosonic string
with $c{=}-2$ matter}, {\tt hep-th/9512092}.

\bibitem{WittenGR}
E~Witten, {\em Ground ring of two dimensional string theory\/},
{\tt hep-th/9108004}, \NPB{373}{1992}{187-213}.

\bibitem{WZ}
E~Witten and B~Zwiebach, {\em Algebraic structures and differential
geometry in 2d string theory\/}, {\tt hep-th/9201056},
\NPB{377}{1992}{55-112}.

\bibitem{Witten}
E~Witten, {\em Topological Quantum Field Theory},
\CMP{117}{1988}{353}.

\end{thebibliography}
\end{document}